\documentclass[conference]{IEEEtran}
\usepackage{booktabs} %
\usepackage{geometry} %
\usepackage{threeparttable}
\ifCLASSINFOpdf
\usepackage{graphicx} 
\else
\fi
\usepackage{multicol}
\usepackage{lipsum}
\usepackage{xcolor}
\usepackage{multirow}
\usepackage{enumitem}
\usepackage{geometry}

\usepackage{fancyhdr}
\newcommand\blfootnote[1]{%
	\begingroup
	\renewcommand\thefootnote{}\footnote{#1}%
	\addtocounter{footnote}{-1}%
	\endgroup
}

\makeatletter
\def\footnoterule{\kern-3\p@
	\hrule \@width 0.49\textwidth \kern 2.6\p@}
\makeatother

\usepackage{caption}
\newcommand{\subparagraph}{}
\usepackage{titlesec}
\titlespacing{\section}
{0pt}{5pt}{5pt}
\titlespacing{\subsection}
{0pt}{5pt}{5pt}

\begin{document}

\begin{flushleft}
	\fontsize{14}{15} \fontfamily{phv}\selectfont
	 \textbf{RT-Focuser: A Real-Time Lightweight Model for Edge-side Image Deblurring}
\end{flushleft}

\begin{flushleft}	
	{\fontsize{10}{10} \fontfamily{phv}\selectfont 
		Zhuoyu Wu\textsuperscript{1,2,3},
        Wenhui Ou\textsuperscript{4}, 
        Qiawei Zheng\textsuperscript{1},
        Jiayan Yang\textsuperscript{1},
        Quanjun Wang\textsuperscript{1},
        Wenqi Fang\textsuperscript{1,2},
        Zheng Wang\textsuperscript{1,2},
        Yongkui Yang\textsuperscript{1,2},
        Heshan Li\textsuperscript{5}
        \\
		\textsuperscript{1}Shenzhen Institutes of Advanced Technology, Chinese Academy of Sciences, Shenzhen, P.R.China\\
		\textsuperscript{2}Shenzhen Unifyware Co., Ltd., Shenzhen, P.R.China\\
        \textsuperscript{3}School of Information Technology, Monash University, Malaysia Campus, Subang Jaya, Malaysia\\
        \textsuperscript{4}Department of Electronic and Computer Engineering, Hong Kong University of Science and Technology, Hong Kong, P.R.China\\
        \textsuperscript{5}Shenzhen Infynova Co., Ltd., Shenzhen, P.R.China
	}
	\blfootnote{\textcolor[rgb]{0,0,0}{This work was funded by National Science
Foundation of China (NSFC) under Grant No.12401676 and No.62372442. (Wenqi Fang and Zheng Wang are the corresponding authors (wq.fang@siat.ac.cn, zheng.wang@siat.ac.cn).)}}

\end{flushleft}

\begin{abstract}
Motion blur caused by camera or object movement severely degrades image quality and poses challenges for real-time applications such as autonomous driving, UAV perception, and medical imaging. In this paper, a lightweight U-shaped network tailored for real-time deblurring is presented and named RT-Focuser. To balance speed and accuracy, we design three key components: Lightweight Deblurring Block (LD) for edge‑aware feature extraction,  Multi-Level Integrated Aggregation module (MLIA) for encoder integration, and Cross-source Fusion Block (X-Fuse) for progressive decoder refinement. Trained on a single blurred input, RT-Focuser achieves 30.67 dB PSNR with only 5.85M parameters and 15.76 GMACs. It runs 6ms per frame on GPU and mobile, exceeds 140 FPS on both, showing strong potential for deployment on the edge. 
The official code and usage are available on: https://github.com/ReaganWu/RT-Focuser.
\end{abstract}

\noindent

\section{Introduction}
\label{Sec: Intro}
Motion blur from camera or object movement degrades visual quality and impairs tasks like autonomous driving, UAV perception, and medical endoscopy. While deep learning methods using CNNs or Transformers show promising results~\cite{cho2021rethinking, zhu2023learning}, their large size and high latency hinder real-time deployment on edge devices. 

Recent works adopt U-shaped networks with multiple inputs and outputs~\cite{nah2017deep, wang2019edvr, wu2024harmonizing}, or event-based methods~\cite{zhu2023learning}, but often suffer from redundancy, complexity, or high per-frame latency ($>$100ms). That impeded the real-time usage in real-time streaming processing. 

To address these issues, we propose \textbf{RT-Focuser}, a Single-Input-Single-Output (SISO) U-shaped network tailored for real-time deblurring. It features: (1) a Lightweight Deblurring Block (LD) with sharpness normalization (SN) to enhance edge preservation; (2) a Multi-Level Integrated Aggregation module (MLIA) to aggregate encoder features; and (3) a Cross-source Fusion Block (X-Fuse) for detail refinement in the decoder.

RT-Focuser offers a strong balance between speed and quality with just 5.85M parameters and 15.76 GMACs. It runs at 6ms/frame on RTX 3090 and 146 FPS on iPhone 15, showing promise for real-time edge deployment.

\begin{figure}[t] 
\centering
\includegraphics[width=\columnwidth]{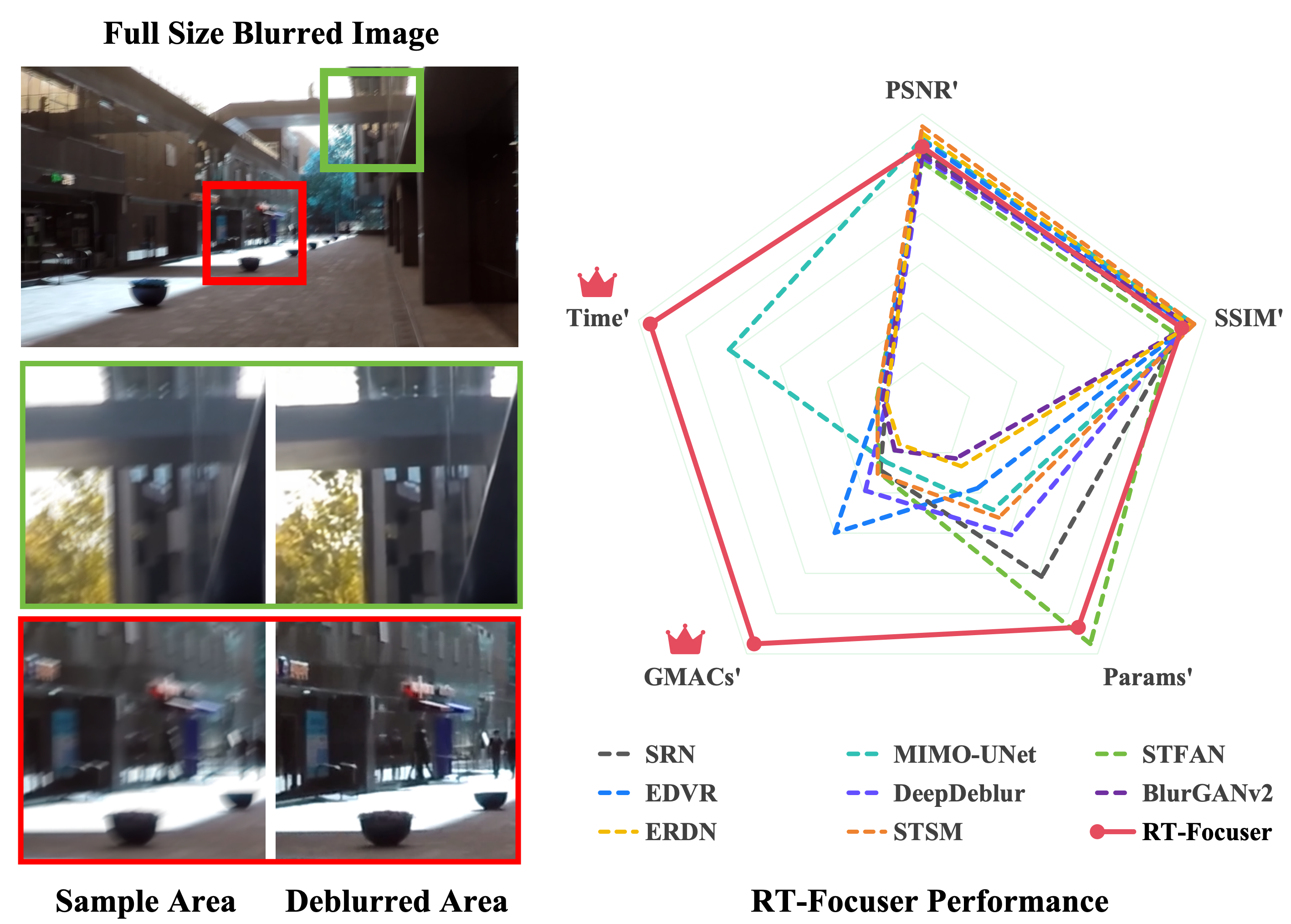} 
\caption{Visual and performance comparison. RT-Focuser shows clear visual recovery (left) and achieves strong trade-offs in PSNR, SSIM, and efficiency metrics (right).}
\vspace{-1mm}
\label{Fig: Intro_Performance}
\end{figure}

\vspace{-1mm}

\section{Methodology}
\label{sec:method}

\begin{figure*}[t]
\centering
\includegraphics[width=\textwidth]{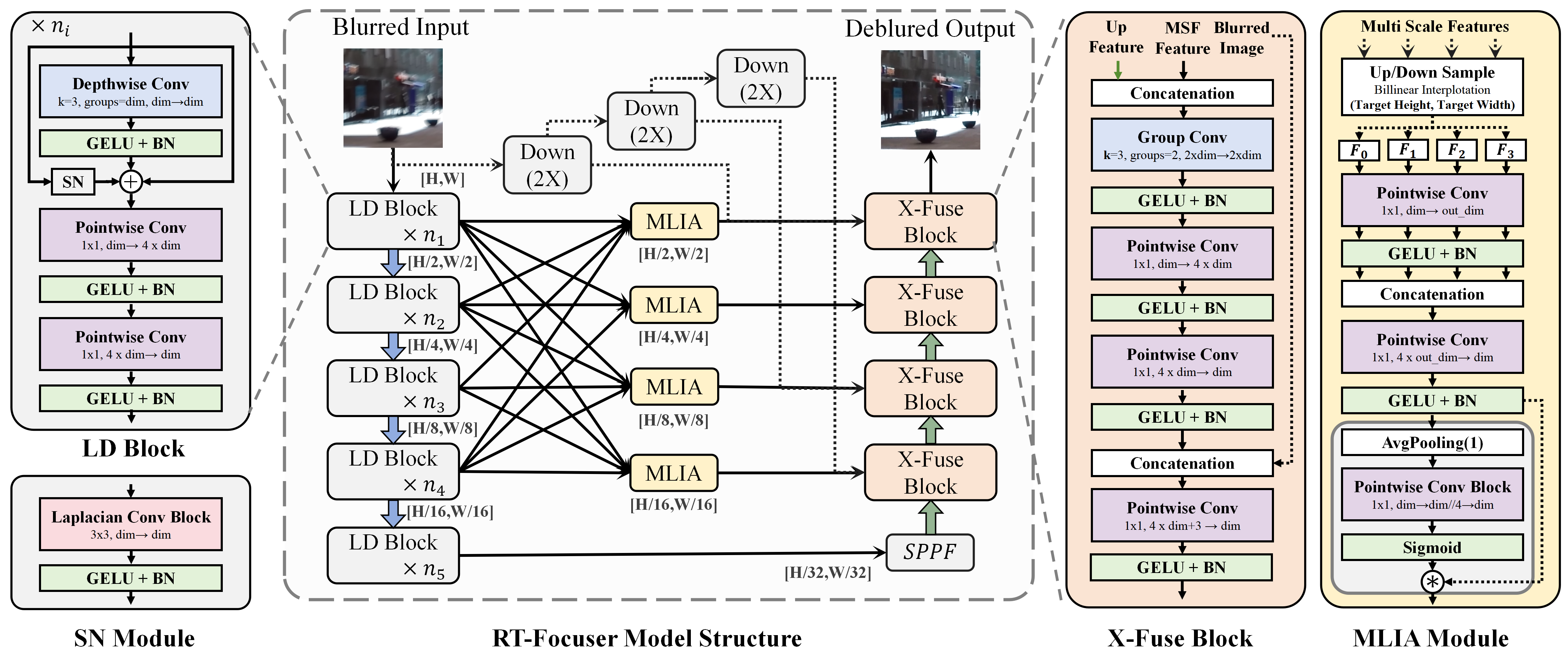}
\caption{Overview of \textbf{RT-Focuser}. It follows a U-shaped encoder–decoder structure with multi-scale fusion. Blue and green arrows indicate 3×3 convolution and ×2 bilinear upsampling, respectively. Down (2X) is the bilinear interpolation for downsampling.}
\label{Fig: Method_Overall}
\end{figure*}

We propose \textbf{RT-Focuser}, a lightweight U-shaped architecture tailored for real-time image deblurring, as illustrated in Fig.~\ref{Fig: Method_Overall}. The network comprises an encoder built with stacked \textbf{LD Blocks}, a decoder with \textbf{X-Fuse Blocks} for progressive refinement, and \textbf{MLIA} module to integrate hierarchical features. 
Additionally, \textbf{SPPF} module is adopted to enhance the receptive field, following the design in YOLO series~\cite{hussain2024yolov5yolov8yolov10goto}.

\subsection{Lightweight Deblurring Block (LD)}
\vspace{-1mm}
LD is designed for efficient extraction. It contains:
\begin{itemize}
  \item A depthwise 3×3 convolution (grouped by dim), followed by GELU and BN;
  \item Two pointwise (1×1) convolutions for channel expansion and compression: dim → 4×dim → dim;
  \item A residual connection optionally enhanced by a Laplacian branch (SN module) to strengthen edge details.
\end{itemize}
\vspace{-2mm}

\subsection{Cross-source Fusion Block (X-Fuse)}
\vspace{-1mm}
At each scale of decoder, the X-Fuse fuses:
\begin{itemize}
  \item Receive inputs: Upsampled features from the previous layer; MSF output from the encoder; The original blurred input (for guidance);
  \item Group and Pointwise convolutions enhance and fuse the inputs in channel and spatial-wise;
  \item Blurry Image is residual concatenated in channel-wise and fused before output.
\end{itemize}
\vspace{-1mm}

\subsection{Multi-Level Integrated Aggregation Module (MLIA)}
\vspace{-1mm}
MLIA aggregates features across encoder stages:
\begin{itemize}
  \item All inputs are resized via bilinear interpolation in shared resolution;
  \item Pointwise convolutions normalize each scale, followed by channel-wise concatenation;
  \item A final 1×1 convolution reduces dimensionality;
  \item An attention branch refines features using global average pooling and a sigmoid gate.
\end{itemize}
\vspace{-1mm}

\section{Experiments}
RT-Focuser is trained for 3000 epochs using AdamW ($lr=1\!\times\!10^{-4}$) with CosineAnnealing, and the loss function is MSE Loss. Experiments are conducted on an RTX 3090 GPU and Xeon 4214R CPU. The GoPro dataset~\cite{nah2017deep} (2,103/1,111 split) is used, with a random $256\!\times\!256$ crop.

\subsection{Comparison with Advanced Models}
We compare RT-Focuser with representative deblurring models in terms of image quality (PSNR, SSIM), complexity (Params, GMACs), and inference latency. As shown in Table~\ref{tab:full_analysis}, RT-Focuser achieves 30.67~dB PSNR with the second lowest parameter count (5.85M), the lowest computation (15.76 GMACs), and the fastest runtime (0.006s). Visual comparisons and sample outputs are presented in Fig.~\ref{Fig: Intro_Performance}.

Compared to large models such as ERDN and BlurGANv2, RT-Focuser reduces computational costs and speeds by more than $100\times$ while maintaining comparable visual quality. Its efficiency and compactness make it ideal for real-time applications.
\vspace{-2.5mm}

\begin{table}[htbp]
\centering
\caption{Comprehensive Model Analysis across Image Quality, Efficiency, and Complexity Metrics}
\label{tab:full_analysis}
\scriptsize
\begin{tabular}{@{}l|ccccc@{}}
\toprule
\textbf{Model} & \textbf{PSNR$\uparrow$} & \textbf{SSIM$\uparrow$} & \textbf{Params$\downarrow$} & \textbf{GMACs$\downarrow$} & \textbf{Time (s)$\downarrow$} \\
\midrule
SRN\cite{tao2018scale}            & 29.97 & 0.9013 & 8.06   & 109.07  & 2.52 \\
MIMO-UNet\cite{cho2021rethinking}      & 31.73 & 0.9500 & 16.10   & 154.41  & 0.014 \\
STFAN\cite{zhou2019spatio}          & 28.59 & 0.8611 & \textbf{5.37}   & 101.18 & 0.15 \\
EDVR\cite{wang2019edvr}           & 31.54 & 0.9260 & 23.61  & 33.44  & 0.21 \\
DeepDeblur\cite{nah2017deep}     & 29.23 & 0.9160 & 11.70  & 62.85  & 4.33 \\
BlurGANv2\cite{kupyn2019deblurgan}      & 29.55 & 0.9340 & 68.20  & 411.34  & 0.35 \\
ERDN\cite{jiang2022erdn}           & 32.48 & 0.9329 & 45.68 & 2138.89 & 2.89 \\
STSM\cite{zhu2023learning}           & \textbf{33.41} & \textbf{0.9512} & 14.40  & 92.51  & 0.16 \\
RT-Focuser     & 30.67 & 0.9005 & 5.85 & \textbf{15.76} & \textbf{0.006} \\
\bottomrule
\end{tabular}
\vspace{3pt}
\begin{tablenotes}
\footnotesize
\item \textbf{Note:} $\uparrow$ indicates higher is better; $\downarrow$ indicates lower is better. PSNR and SSIM reflect image quality. Params, GMACs, and runtime (per image) represent model complexity and inference efficiency.
\end{tablenotes}
\end{table}
\vspace{-1.5mm}

\subsection{Deployment Efficiency on Edge and General Platforms}

To assess real-time performance, we benchmark RT-Focuser on four platforms: GPU, mobile SoC, and CPU with different backends. As shown in Table~\ref{tab:deployment}, it achieves over 140 FPS on GPU and mobile, and maintains reasonable speed on CPUs.

\begin{table}[h]
\centering
\caption{RT-Focuser Deployment Speed on Different Platforms}
\label{tab:deployment}
\scriptsize
\begin{tabular}{@{}l|c|l@{}}
\toprule
\textbf{Platform} & \textbf{FPS$\uparrow$} & \textbf{Backend Details} \\
\midrule
iPhone 15 (A16 Bionic) & \textbf{146.72} & CoreML \\
RTX 3090 GPU & \textbf{154.42} & PyTorch CUDA \\
Intel CPU (Xeon) & 14.95 & ONNX Runtime \\
Intel CPU (Xeon) & 22.74 & OpenVINO \\
\bottomrule
\end{tabular}
\vspace{3pt}
\begin{tablenotes}
\footnotesize
\item \textbf{Note:} FPS measured for $256\times256$ input size. All measurements use batch size 1 and single-thread inference unless otherwise specified.
\end{tablenotes}
\vspace{-3mm}
\end{table}

\section{Conclusion}
We present \textbf{RT-Focuser}, a lightweight and efficient network for real-time image deblurring. Through the design of the LD Block, MLIA module, and X-Fuse block, RT-Focuser achieves a strong balance between restoration quality and computational cost. RT-Focuser outperforms existing lightweight models in both speed and parameter efficiency, while maintaining competitive PSNR and SSIM. Moreover, the model achieves over \textbf{140 FPS} on mobile and GPU platforms, highlighting its practicality for real-time deployment in Edge-Side.

\ifCLASSOPTIONcaptionsoff
  \newpage
\fi

\bibliographystyle{IEEEtran}  %
\bibliography{refs}  

\end{document}